\documentclass[10pt,aps,pr,twocolumn,showpacs,amssymb,floatfix]{revtex4-1}
\usepackage{amsmath,graphics,epsfig,mathrsfs,physics}
\usepackage{bm}
\usepackage{epigraph}

\usepackage{color}

\newcommand{\etal}{{\em et al.~}}

\usepackage{color}
\definecolor{darkgreen}{rgb}{0.0, 0.5, 0.0}
\definecolor{brown}{rgb}{0.59, 0.29, 0.0}
\definecolor{darkgreen2}{rgb}{0.01, 0.75, 0.24}
\usepackage{hyperref}
\hypersetup{backref=true,pagebackref=true,hyperindex=true,colorlinks=true,citecolor=blue, breaklinks=true,urlcolor=
blue,linkcolor=blue,bookmarks=true,bookmarksopen=true}

\begin{document}
\title{Dirac Spin-Orbit Torques and Charge Pumping at the Surface of Topological Insulators}
\author{Papa B. Ndiaye$^{1}$}
\email{papabirame.ndiaye@kaust.edu.sa}
\author{C. A. Akosa$^{1}$}
\author{M. H. Fischer$^{2}$}
\author{A. Vaezi$^{3,4}$}
\author{E-A. Kim$^{3}$}
\author{A. Manchon$^{1,5}$}
\email{aurelien.manchon@kaust.edu.sa}
\affiliation{$^1$King Abdullah University of Science and Technology (KAUST), Physical Science and Engineering Division (PSE), Thuwal 23955-6900, Saudi Arabia.}
\affiliation{$^2$Institute for Theoretical Physics, ETH Zurich, 8093 Zurich, Switzerland.}
\affiliation{$^3$Department of Physics, Cornell University, Ithaca, New York 14853, USA.}
\affiliation{$^4$Department of Physics, Stanford University, Stanford, CA 94305, USA}
\affiliation{$^5$King Abdullah University of Science and Technology (KAUST), Computer, Electrical and Mathematical Science and Engineering Division (CEMSE), Thuwal 23955-6900, Saudi Arabia.}

\begin{abstract}        
We address the nature of spin-orbit torques at the magnetic surfaces of topological insulators using the linear response theory. We find that the so-called Dirac torques in such systems possess a different symmetry compared to their Rashba counterpart, as well as a high anisotropy as a function of the magnetization direction. In particular, the damping torque vanishes when the magnetization lies in the plane of the topological insulator surface. We also show that the Onsager reciprocal of the spin-orbit torque, the charge pumping, induces an enhanced anisotropic damping. Via a macrospin model, we numerically demonstrate that these features have important consequences in terms of magnetization switching.
\end{abstract}
\pacs{75.60.Jk,85.75.Dd,72.25.-b}
\maketitle

\section{Introduction}
Not only has spintronics yielded to the market real deal solutions for low energy, high density non-volatile memory \cite{chappert2007} but it has also provided a fundamental understanding of the different mechanisms by which efficient electrical control of spin currents and magnetic configurations are possible. The spin transfer torque (STT) mechanism \cite{stt}, which is central to a whole generation of memory devices, exploits the transfer of spin angular momentum between a spin current flow and the local magnetization of a ferromagnetic (FM) layer thereby enabling magnetization switching or precession \cite{sttexp}. A critical hurdle for traditional STT setups is the need for a spin-polarizer generating the spin-current: STT devices comprise a number of ultrathin (anti-)ferromagnetic, metallic and insulating layers (see e.g. Ref. \cite{Yuasa2013}), rendering the design of architectures rather complex.

Research to circumvent this issue and enhance the efficiency of torque generation led to the proposal of the spin-orbit torques (SOTs) \cite{sot2009,garate2009}, which arise from the transfer of angular momentum between a flowing charge current and the local magnetization mediated by spin-orbit coupling. Systems with inversion symmetry breaking, such as magnetic multilayers involving heavy metals (HM), are  excellent platforms for the realization of magnetization reversal induced by in-plane charge currents \cite{miron2010,sheliu2012}: the HM provides a large spin Hall effect (SHE) while the FM/HM interface supports sizable Rashba spin-orbit coupling  \cite{manchon2015}, both at the origin of large SOTs \cite{haney,rashba,dft}. Various features have been identified experimentally, such as large angular anisotropies \cite{garello} and complex materials dependence \cite{masa}. Innovative concepts such as spin wave mediated \cite{manchon2014} or intrinsic SOTs have also been introduced \cite{kurebayashi,Li2015}, thereby broadening the field of spin-orbitronics which opens new roads for spin devices made of engineered (non)magnetic materials and operated without magnetic fields \cite{Ndiaye2017}.

The recent observation of SOTs in magnetic bilayers involving topological insulators (TI) offers an alternative route towards efficient electrical control of the magnetization \cite{mellnik2014,fan2014}. A three dimensional (3D) TI is topologically distinct from a conventional 3D band insulator: it possesses an insulating bulk while hosting chiral metallic channels at the edges, where electrons are described as massless Dirac fermions with tight interlock between spin and momentum \cite{hasankane2010}. The strong spin-momentum locking results in large spin-charge conversion efficiency \cite{Shiomi2014,deorani14,Jamali2015,Rojas2016}, as well as large SOTs enabling the control of adjacent magnetic layers \cite{Fan2016,Wang2015}. The main strategies adopted so far consist in either doping the TI with magnetic impurities \cite{chen2010,fan2014} or using the proximity effect by coating it with (possibly insulating) ferromagnets  \cite{scholz2012,mellnik2014}. \par
Various phenomena such as the topological magnetoelectric effect \cite{Qi2008,garate2010}, STT and current-driven magnetization dynamics \cite{yokoyama2010,yokoyama2011,nomura2010,tserk2012,tserk2015,linder14}, the interplay between spin and charge \cite{burkov2010,Ueda2012,liu2013,chang2015,Okuma2016}, as well as spin transport in magnetic TIs \cite{Fujimoto2014,sakai2014,fischer2016,taguchi2015,mahfouzi2014,mahfouzi2016} have been studied theoretically. Despite these important theoretical efforts, major puzzles remain to be understood such as the emergence of gigantic damping-like torque \cite{mellnik2014,fan2014}, the sizable angular dependence of the SOT \cite{fan2014} and the significant discrepancies between the spin-charge conversion rates reported in the SOT experiments \cite{mellnik2014,fan2014,Fan2016,Wang2015} and the spin pumping experiments \cite{Shiomi2014,deorani14,Jamali2015,Rojas2016}. It is still unclear whether the spin-charge conversion efficiencies reported experimentally can be solely attributed to topological surfaces states \cite{Zhang2016}. It is therefore crucial to establish a solid understanding of the physics at stake at the magnetic surface of TIs in order to properly interpret these experimental results. \par


\begin{figure}[b]
\includegraphics[width=8.4cm]{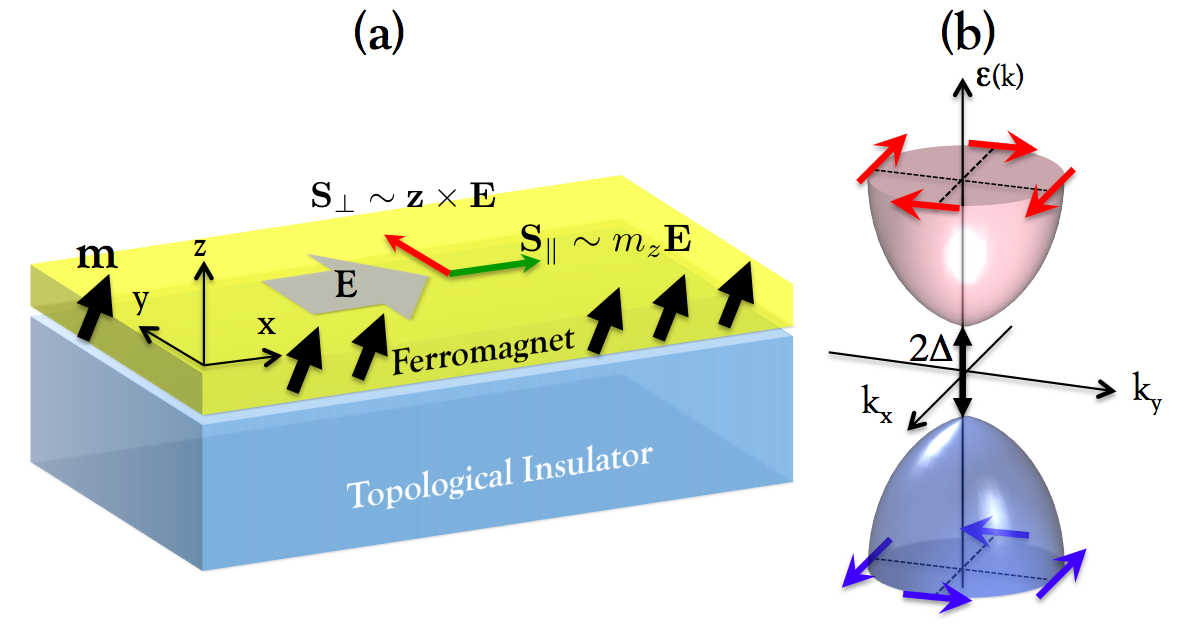}
\caption{\label{fig:1} (Color online) (a) Bilayer consisting of a TI substrate with a ferromagnetic overlayer. The black arrows represent the local magnetic moment with overall magnetization direction ${\bf m}$. The electric field ${\bf E}$ is applied along ${\bf x}$ and generates two spin density components, ${\bf S}_\bot\sim {\bf z}\times{\bf E}$ (red arrow) and ${\bf S}_\|\sim m_z{\bf E}$ (green arrow). (b) Schematics of the two dimensional band structure at the magnetic surface of the TI when ${\bf m}={\bf z}$. The red (blue) arrows represent the spin direction in the conduction (valence) band and the perpendicular magnetization opens a gap $2\Delta$.} 
\end{figure}

In this work, we explore the nature and the symmetry of nonequilibrium spin densities, their coupling to the magnetic order at TI magnetic surfaces, and discuss their differences with respect to other spin-orbit generated spin densities via spin Hall effect \cite{haney} and Rashba effect \cite{sot2009,rashba}. In Section II, we present the model and address the electrically driven Dirac SOTs in magnetized topological insulators using  the Kubo formula within the linear response theory. We show that the effective Dirac SOT is of the form ${\bf T}= T_\|(m_z) m_z {\bf m}\times{\bf E}$ + $T_\bot(m_z){\bf m}\times ({\bf z}\times{\bf E})$, where the in-plane and out-of-plane torques $T_{\|,\bot}(m_z)$ exhibit a sizable anisotropy ($m_z$ is the projection of the magnetization ${\bf m}$ to the normal  ${\bf z}$ to the TI surface and ${\bf E}$ is the applied electric field). In Section III, we discuss the reciprocal effect, i.e. the charge current pumped by magnetization dynamics and show that it produces an enhanced anisotropic magnetic damping torque. Finally in Section IV, we demonstrate numerically using the Landau-Lifshitz-Gilbert equation that the Dirac torque can reverse the magnetization in layers with perpendicular magnetic anisotropy, but is formally less efficient than the torque arising from spin Hall effect. 

\section{Electrically driven Dirac SOTs}

Let us start by considering the top surface of a three-dimensional TI in the presence of magnetic exchange, as depicted in Fig. \ref{fig:1}. Near the Dirac point, the simplest low-energy effective Hamiltonian of the conducting surface states reads
\begin{eqnarray} \label{H}
\hat H&=&\hat H_0+\hat H_i,\\\label{H0}
\hat H_0&=&\hbar v\hat{\boldsymbol{\sigma}}\cdot({\bf k}\times{\bf z})+\Delta \hat{\boldsymbol{\sigma}}\cdot{\bf m}-\varepsilon_{\rm F},\\\label{Hi}
\hat H_i&=&\sum_i V_0 \delta ({\bf r}-{\bf r}_i),
\end{eqnarray}
where $\hat H_0$ is the translationally invariant and time-independent unperturbed Hamiltonian and $\hat H_i$ accounts for random short-range disorder, treated as a perturbation in this work. In Eq. \eqref{H0}, the first term is the usual Rashba-type spin-orbit coupling with $v$ the Fermi velocity ($\simeq6\times10^5$ m.s$^{-1}$ in Bi$_2$Se$_3$ and $4.3\times10^5$ m.s$^{-1}$ in Bi$_2$Te$_3$). The electron transport is confined in the ($x,y$) plane and ${\bf k} =(k_x,k_y,0)=k(\cos\phi_k,\sin\phi_k,0)$. The second term in Eq. \eqref{H0} is the exchange coupling between itinerant and local spins. Here, $\hat{\boldsymbol{\sigma}}$ is the vector of Pauli matrices and $\Delta$ is the exchange energy and the magnetization ${\bf m}=(m_x,m_y,m_z)=(\sin\theta\cos\phi,\sin\theta\sin\phi,\cos\theta)$ is uniform and can point along any (general) direction. The last term is the Fermi energy, emphasizing that we are interested in the metallic regime where the chemical potential stands away from the charge neutrality point.\par

The out-of-plane magnetization component is responsible for the gap opening in the TI spectrum via $\Delta m_z \hat\sigma_z$, thereby providing the mass of Dirac fermions. Indeed, the  $unperturbed$ Hamiltonian $\hat H_0$  can be re-written as
\begin{align}\label{H0b}
\hat{H}_0
= \hbar v({\bf z}\times\hat{\boldsymbol{\sigma}})\cdot\left( {\bf k} +e{\bf A}\right) +\Delta m_z \hat\sigma_z-\varepsilon_{\rm F},
\end{align}
where $e{\bf A}=\frac{\Delta}{\hbar v}{\bf z}\times\bf m$ is identified as the effective vector potential \cite{nomura2010}. Hence, the $m_{x,y}$ components of the magnetization do not open a gap in the energy dispersion but only shift the Dirac cone along the $k_{x,y}$ - direction. These in-plane magnetization components are not expected to impact any physical observables as they can be straightforwardly removed by redefining the position of the Dirac node. Another particular feature of this model is that the velocity operator $\hat{\bf v}=\partial_{\hbar{\bf k}}\hat{H}_0 $ is indeed directly proportional to the spin operator as $\hat{\bf v}= v({\bf z}\times\hat{\boldsymbol{ \sigma}})$, drawing an equivalence between the electric current $\bf j$ at the surface of magnetic TIs and the in-plane components of the spin density $\bf S$  \cite{raghu2010,tserk2015,sakai2014,fischer2016},
\begin{align}\label{eq:js}
{\bf j}=-ev {\bf z}\times{\bf S}.
\end{align}
This spin-velocity identity in TIs is echoed in the expressions of the response functions such as the conductivity tensor characterizing the electrical transport and the dynamical spin susceptibility. Therefore, the coefficients of the damping-like and field-like torques derived below correspond to the diagonal and off-diagonal Hall conductivities, respectively, due to the spin-momentum lock. The anomalous Hall conductivity in particular has been analyzed by others in Dirac systems \cite{sinitsyn2006,sinitsyn2007,Xiao2007,Ezawa2012,ado2015} and Weyl semimetals \cite{nomura2015,burkov2014}.\par

The chiral basis eigenstates that diagonalize the unperturbed Dirac Hamiltonian $\hat{H}_0$ are explicitly written as 
\begin{eqnarray}\label{eq:chiralbasis}
\ket{u_+^{\bf k}}=\begin{pmatrix}e^{i\gamma_k}\cos\frac{\chi_k}{2}\\ \sin\frac{\chi_k}{2}\end{pmatrix},\;\ket{u_{-}^{\bf k}}=\begin{pmatrix}-e^{i\gamma_k}\sin\frac{\chi_k}{2}\\\cos\frac{\chi_k}{2}\end{pmatrix},
\end{eqnarray}
with
$$\tan\gamma_k=\frac{\hbar vk\cos\phi_k-\Delta\sin\phi\sin\theta}{\hbar vk\sin\phi_k+\Delta\cos\phi\sin\theta},\;\cos\chi_k=\frac{\Delta}{|\varepsilon_{\bf k}^s|}\cos\theta,$$ and $\varepsilon_{k}^{s}=s\sqrt{\hbar^2v^2k^2+\Delta^2+2\hbar vk\Delta\sin\theta\sin (\phi_k-\phi)}$. The expectation value of the spin density for state $s$ is therefore 
\begin{eqnarray}
\langle{\bf S}\rangle_s=\frac{\Delta}{\varepsilon_{\bf k}^s}{\bf m}-\frac{\hbar v}{\varepsilon_{\bf k}^s}{\bf z}\times{\bf k}.
\end{eqnarray}
The expectation value of the spin density contains two distinct contributions, a component aligned with the magnetization $\sim{\bf m}$ and an in-plane component proportional to $\sim{\bf z}\times{\bf k}$. Only the latter produces a non-equilibrium spin density and therefore, one should not expect a current-driven $S_z$ component.

Let us first express the electrically-driven nonequilibrium spin density $\delta{\bf S}$ in the framework of linear response theory. The Streda-Smrcka version \cite{streda} of the Kubo formula yields two contributions \cite{sinitsyn2006}
\begin{multline} \label{kubostreda}
\delta {\bf S}^{\rm I}=\frac{\hbar}{2\pi{\cal V}}{\rm Re}\int_{-\infty}^{+\infty} d\varepsilon \partial_\varepsilon f(\varepsilon) {\rm tr}
\Big[\hat{\bm{\sigma}}\hat G^A_{\varepsilon}(\hat{\bf v}\cdot e {\bf E})(\hat G^R_{\varepsilon}-\hat G^A_{\varepsilon}) \Big],
\end{multline}
\begin{multline} \label{kubostreda1}
\delta {\bf S}^{\rm II}=\frac{\hbar}{2\pi{\cal V}}{\rm Re}\int_{-\infty}^{+\infty} d\varepsilon  f(\varepsilon){\rm tr}\Big[
\hat{\bm{\sigma}}\hat G^R_{\varepsilon}(\hat{\bf v}\cdot e {\bf E})\partial_\varepsilon\hat G^R_{\varepsilon}\\
-\hat{\bm{\sigma}}\partial_\varepsilon\hat G^R_{\varepsilon}(\hat{\bf v}\cdot e {\bf E})\hat G^R_{\varepsilon}\Big].
\end{multline}
$\hat G_\varepsilon^\alpha$ are the Green's functions defined in momentum and energy space, ${\cal V}$ is the volume of the unit cell and ${\rm tr}$ accounts for the trace on the spin space as well as the summation over the ${\bf k}$-space.
The Fermi-surface contribution $\delta {\bf S}^{\rm I} \propto \partial_\varepsilon f(\varepsilon)$ is complemented by the Fermi-sea contribution $\delta {\bf S}^{\rm II}$. However, in the case of any two-dimensional Dirac model such as the TI surface considered here, this second contribution vanishes in the metallic regime, i.e. $\varepsilon_{\rm F}>\Delta>0$ (see Appendix for details). The weak impurities we consider here not only broaden the energy levels by introducing a finite lifetime $\tau$ to the quasiparticle in the chiral bands, but also change the eigenstates. The quasiparticle lifetime broadening induced by the presence of impurities is reflected in the retarded self-energy, which is defined self-consistently as \cite{sakai2014}
\begin{equation}
\hat\Sigma^R=n_i V_0^2 \int \frac{d^2{\bf k}}{(2\pi)^2} \left[\hat G_0^R+ \partial_k\hat G^R(\hat\Sigma^R)\right].
\end{equation} 
For Dirac electrons, the calculation of the entire retarded self-energy in an environment with $\delta$-type impurities has to be done with care as logarithmic divergences naturally occur \cite{shon1998}. The first term of the self-energy is diagonal (only with $\hat{\sigma}_0$ and $\hat{\sigma}_z$ components), $\bf k$-independent and readily writes as 
\begin{equation}
\hat\Sigma_{\parallel}^R=- \frac{i\hbar}{4\tau}(1+\beta m_z\hat{\sigma}_z),
\end{equation} 
where the impurity scattering rate is given by $\hbar/\tau=n_i V_0^2 k_{\rm F}/\hbar v$ and $\beta=\Delta/\varepsilon_{\rm F}$ is the spin polarization. The second term is $\bf k$-dependent and off-diagonal ($\equiv \Sigma_{\rm x}\hat{\sigma}_x+\Sigma_{\rm y}\hat{\sigma}_y$), and the $\bf k$-integration should be done here with the impurity-range ultraviolet cutoff \cite{fujimoto2013} in order to respect gauge invariance via the Takahashi-Ward identity \cite{mahan}. The detailed procedure described in Refs. \cite{fujimoto2013,sakai2014} leads to the renormalization of the velocity of the Dirac electron as $\tilde v=(1-\xi)v$ with $\xi=\frac{n_i V_0^2}{4\hbar^2v^2} \ll 1$ within the weak impurity limit and the full renormalized retarded self-energy reading now as $\hat\Sigma^R=\hat\Sigma_{\parallel}^R+\hbar\xi \tilde v[(\hat{\boldsymbol{\sigma}}\times{\bf k})\cdot{\bf z}]$. In the self-consistent Born approximation, the retarded Green's function reads
\begin{equation} \label{scbaGF}
\hat G_\varepsilon^R=\frac{\varepsilon+[\hbar\tilde{v}({\bf z}\times {\bf k})+\Delta {\bf m}]\cdot\hat{\boldsymbol{\sigma}}+\frac{i\hbar}{4\tau}(1-\beta m_z \hat\sigma_z)}{(\varepsilon-\varepsilon_{\bf k}^{+}+i\Gamma^{+})(\varepsilon-\varepsilon_{\bf k}^{-}+i\Gamma^{-})},
\end{equation}
where $\Gamma^{\pm}=\frac{\hbar}{4\tau}(1\pm \beta^2m_z^2)$. Inserting the perturbed Green's function, Eq. \eqref{scbaGF}, into Eq. \eqref{kubostreda} is not sufficient to fully capture the impact of the impurities. As a matter of fact, the proper calculation of $\delta \bf S$ includes a variety of crossing and non-crossing diagrams of the same order which have to be properly accounted for \cite{mahan,ado2015,ado2016}. We adopt here the common approximation by selecting only the non-crossing ladder diagrams through the so-called vertex correction to the spin operator \cite{rammerbook,mahan}. Notice that it has been shown recently that a more accurate evaluation should include a subclass of crossing diagrams in addition to the standard set of non-crossing ones \cite{ado2015,ado2016}. However, since we assume a low impurity concentration, the average distance between the impurities is larger than the Fermi wavelength of the electronic carriers and we limit ourselves to the ladder non-crossing approximation. The spin operator $\hat{\bm\sigma}$ in Eq. \eqref{kubostreda} is then replaced by a renormalized operator $\hat{\bm \Upsilon}$ that must satisfy $\hat\Upsilon_i=\hat{\sigma}_i+n_iV_0^2\int\frac{d^2{\bf k}}{(2\pi)^2}\hat{G}^R\hat\Upsilon_i\hat{G}^A$ \cite{sinitsyn2006}. By writing $\hat{ \Upsilon}_i$ in the tensor form ($\hat{\Upsilon}_i=\Sigma_j\varsigma_i^j\hat{\sigma}_j$, $j=0,x,y,z$), we find the two components $\hat\Upsilon_x={\cal A}\sigma_x+{\cal B}\sigma_y$ and $\hat\Upsilon_y=-{\cal B}\sigma_x+{\cal A}\sigma_y$,
where ${\cal A}=2\frac{1+\beta^2m_z^2}{1+3\beta^2m_z^2}$ and ${\cal B}=\frac{2\hbar}{\tau\varepsilon_{\rm F}}\frac{\beta m_z(1+\beta^2m_z^2)}{(1+3\beta^2m_z^2)^2}$.
The vertex-corrected version of the spin density $\delta \bf S$ in Eq. \eqref{kubostreda} gives
\begin{align} \label{sden}
\delta{\bf S} =&-\frac{\tau\varepsilon_{\rm F}}{2\hbar \tilde{v}\pi}\frac{1-\beta^2m_z^2}{1+3\beta^2m_z^2}{\bf z}\times e{\bf E} \nonumber\\&-\frac{\beta}{\tilde{v}\pi }\frac{1+\beta^2m_z^2}{(1+3\beta^2m_z^2)^2}m_z e{\bf E}.
\end{align}
First, we emphasize that the non-equilibrium electrically driven spin density is in-plane and does not have any $z$-component, contrary to the Rashba model \cite{rashba,Li2015}. As a matter of fact, in the Dirac model the flowing electrons only experience the out-of-plane component of the magnetization [$\sim \Delta m_z\hat{\sigma}_z$ in Eq. (\ref{H0b})] and therefore their precession about the magnetization direction only mixes $S_x$ and $S_y$ components, but does not yield any $S_z$ component. The Dirac spin-orbit torque, ${\bm\tau}=(2\Delta/\hbar) {\bf m}\times\delta{\bf S}$, straightforwardly yields 
\begin{align}\label{eq:torque}
{\bm\tau} =&-\frac{\beta\tau\varepsilon_{\rm F}^2}{\hbar^2\tilde{v}\pi}\frac{1-\beta^2m_z^2}{1+3\beta^2m_z^2}{\bf m}\times({\bf z}\times e{\bf E}) \nonumber\\&-\frac{2\beta^2\varepsilon_{\rm F}}{\hbar\tilde{v}\pi}\frac{1+\beta^2m_z^2}{(1+3\beta^2m_z^2)^2}m_z {\bf m}\times e{\bf E}.
\end{align}
Notice that this form is more general than the one derived in Ref. \cite{sakai2014} which is restricted to ${\bf m}={\bf z}$. The effective Dirac SOT is found to be of the form ${\bm\tau}= \tau_\| m_z {\bf m}\times e{\bf E} + \tau_\bot {\bf m}\times ({\bf z}\times e{\bf E})$, a form similar to the one obtained in the limit of large Rashba spin-orbit coupling in a magnetic Rashba gas \cite{Li2015}. The first term is odd upon magnetization reversal, proportional to the current flow ($\propto \tau$) and acts like a field-like torque. In contrast, the second term, $\sim m_z{\bf m}\times e{\bf E}$, is even in magnetization reversal, independent of scattering and acts like a damping torque. While the former arises from the traditional inverse spin galvanic effect \cite{isge1990}, the latter is the magnetoelectric coupling identified by Garate and Franz \cite{garate2010}. This damping torque is quite different from the damping-like torque stemming from spin Hall effect, usually observed in magnetic bilayers involving heavy metals \cite{sheliu2012,garello,masa}. Indeed, the magnetoelectric effect at the surface of topological insulators vanishes when the magnetization lies in the plane of the surface \cite{garate2010}, while the SHE-induced damping torque ($\sim {\bf m}\times[({\bf z}\times{\bf E})\times{\bf m}]$) remains finite. Notice also that the sign of the SOT reported in Eq. (\ref{eq:torque}) is opposite to the one derived for the magnetic Rashba gas \cite{Li2015}, which is attributable to the spin chirality of the Dirac conduction band. Due to the identity between the spin and the velocity operators, Eq. (\ref{eq:js}), the field-like and damping-like Dirac torques coefficients correspond to the longitdudinal and transverse (Hall) conductivities, respectively \cite{sinitsyn2006,sinitsyn2007,sakai2014}. Finally, the SOT in Eq. (\ref{eq:torque}) exhibits a complex dependence as a function of the magnetization direction, associated with the distortion of the band structure when the magnetization lies perpendicular to the surface.

\section{Charge pumping and Anisotropic Damping}

While a charge current can exert a torque on the local magnetization, a precessing magnetization can pump a charge current \cite{brataas11,freimuth15}. These two effects are related to each other via Onsager reciprocity relation and can be treated on equal footing. The spin-to-charge conversion process has been investigated experimentally in ferromagnet/topological insulator heterostructures \cite{Shiomi2014,deorani14,Jamali2015,Rojas2016} (e.g. FM/Bi$_2$Se$_3$), and some of its aspects have been treated theoretically \cite{mahfouzi2014,taguchi2015}. \par

The magnetization dynamics under an external magnetic field and SOT is given by the Landau-Lifschifz-Gilbert (LLG) equation
\begin{eqnarray}\label{eq:llg}
\partial_{t} {\bf m}=(\gamma/M_{\rm s}) {\bf m} \times \partial_{{\bf m}}{\cal F} +\hat{\kappa}\cdot {\bf E}.
\end{eqnarray}
Here, $\partial_{{\bf m}}{\cal F}$ is the functional derivative of the magnetic energy density ${\cal F}$ that governs the dynamics of the magnetization in the absence of charge flow while ${\bf E}$ is the electric field that drives the SOT through the tensor ${\hat \kappa}$. $\gamma$ and $M_{\rm s}$ are the absolute value of the gyromagnetic ratio and the saturation magnetization, respectively. The charge current density reads
\begin{eqnarray}\label{eq:charge}
{\bf J}_{c}=\hat{\bf{\delta}} \cdot \partial_{{\bf m}}{\cal F}+\hat{g} \cdot {\bf E},
\end{eqnarray}
where the electric field drives the charge current through the conductivity tensor $\hat{g}$, while the magnetization dynamics pumps a charge current through the tensor ${\hat \delta}$. Let us now consider a magnetic layer of width $w$, thickness $d$, length $L$ and section normal to the current flow $ S = w d$. The particle current is defined as $\partial_{{t}}n_{i}= S {\rm J}_{c,i}/e$ , and the electric and magnetic potentials driving the charge and magnetization dynamics respectively read $f^{j}_{e}=Le E_{j}$, $f^{j}_{m}=\Omega\partial_{m_{j}} {\cal F}$. Here $\Omega=Lwd$ is the volume of the magnet. Therefore, Eqs. (\ref{eq:llg}) and (\ref{eq:charge}) can be rewritten in the more convenient form $\left(                 
  \begin{array}{cc}   
   \partial_{t}n_{i}  \\  
   \partial_{t}m_{i}\\  
  \end{array}
\right)   
= \hat{\cal L}
\left(                 
  \begin{array}{cc}   
   f^{j}_e\\  
   f^{j}_m\\  
  \end{array}
\right)$, where the Onsager coefficients in $\hat{\cal L}$ are explicitly expressed  as
\begin{equation}       
\left(                 
  \begin{array}{cc}  
    {\cal L}_{n_{i},f^{j}_e} & {\cal L}_{n_{i},f^{j}_m} \\ 
    {\cal L}_{m_{i},f^{j}_e} & {\cal L}_{m_{i},f^{j}_m}\\ 
  \end{array}
\right)   
=\left(                 
  \begin{array}{cc}   
   \frac{S}{Le^2}g_{ij} & \frac{1}{Le}{\delta}_{ij} \\  
    \frac{1}{Le}\kappa_{ij} &-\frac{\gamma}{M_{\rm s}\Omega}({\bf e}_{i}\times{\bf e}_{j})\cdot\bf m
  \end{array}
\right).                 
\end{equation}
When applying Onsager reciprocity principle \cite{brataas11,onsager}
\begin{align}
 L_{n_{i},f^{j}_m}({\bf m})=-L_{m_{j},f^{i}_e}(-{\bf m}),
\end{align}
 we get ${ \delta}_{ij}({\rm\bf m})=-{\kappa}_{ji}(-{\rm\bf m})$. In the previous section, Eq. \eqref{eq:torque}, we showed that the torque density $\bm \tau$ at the surface of the TI reads
\begin{align}
{\bm\tau}=\tau_\|m_z{\rm\bf m}\times {\bf E}+\tau_\bot{\rm\bf m}\times({\rm\bf z}\times{\rm\bf E}),
\end{align}
where $\tau_{\|,\bot}$ are the damping-like and field-like components, respectively. The total torque exerted on the ferromagnet is then ${\bf T}=\hat{\kappa}\cdot{\bf E}=\int dA {\bm\tau}$ ($A=Lw$ is the surface area), which yields
 \begin{align}
{ \kappa}_{ij}=\frac{\mu_{\rm B}}{M_{\rm s}d}\left\{\tau_\|m_z({\bf m}\times {\bf e}_{j})\cdot {\bf e}_{i}+\tau_\bot[{\bf m}\times({\bf z}\times{\bf e}_{j})]\cdot {\bf e}_{i} \right\}. \nonumber\\
\end{align}
By direct application of the Onsager reciprocity relation, we then deduce the charge pumping coefficients in TIs 
\begin{align}
{ \delta}_{ij}=\frac{\mu_{\rm B}}{M_{\rm s} d}\left\{-\tau_\|m_z({\bf m}\times {\bf e}_{i})\cdot {\bf e}_{j}+\tau_\bot[{\bf m}\times({\bf z}\times{\bf e}_{i})]\cdot {\bf e}_{j}\right\}. \nonumber\\
\end{align}
The charge current pumped by the magnetization dynamics simply reads
\begin{align}\label{eq:Jcp}
{\bf J}_{c}^{\rm pump}= & \frac{\hbar}{2d}\left(\tau_\|m_z{\partial_{t}\bf m}+\tau_\bot{\bf z}\times\partial_{t}\bf m\right).
\end{align}
This equation establishes the correspondence between the current-driven Dirac SOT and the charge current pumped by a time-varying magnetization. 
By the virtue of Onsager reciprocity, the results and conclusions drawn above for the SOT apply straightforwardly to the charge pumping through Eq. (\ref{eq:Jcp}), in particular the second component $\sim {\bf z}\times\partial_t{\bf m}$ dominates in the metallic regime since $\tau_\bot>\tau_\|$. Notice that ${\bf J}_{c}^{\rm pump}$ is the current density flowing in the magnetic volume and is therefore inversely proportional to the thickness $d$.\par

The charge current pumped at the surface of the TI, $d{\bf J}_c^{\rm pump}$, also induces an interfacial non-equilibrium spin density $\delta{\bf S}_{\rm pump}=(d/e v){\bf z}\times{\bf J}_c^{\rm pump}$ [see Eq. (\ref{eq:js})]. In turn, this pumped interfacial spin density induces a torque, ${\bf T}_{\rm pump}=(2\Delta/\hbar)\int dA{\bf m}\times\delta{\bf S}_{\rm pump}$, that reads
\begin{eqnarray}
{\bf T}_{\rm pump}&=&\frac{\mu_{\rm B}}{M_{\rm s} d}\frac{\Delta}{e v}\tau_\bot{\bf m}\times[{\bf z}\times(\partial_t{\bf m}\times{\bf z})]\nonumber\\
&&+\frac{\mu_{\rm B}}{M_{\rm s} d}\frac{\Delta}{e v}\tau_\|m_z^2{\bf z}\times(\partial_t{\bf m}\times{\bf z}).
\end{eqnarray}
The first term is {\em odd} upon time reversal operation ($\partial_t\rightarrow-\partial_t$, ${\bf m}\rightarrow-{\bf m}$), while the second term is {\em even}. Accordingly, the first term contributes to the magnetic damping, while the second term renormalized the gyromagnetic ratio. In particular, the damping torque acts only on the {\em in-plane} components of the magnetization ($m_x,m_y$), thereby creating an anisotropic damping. The total magnetic damping then reads
\begin{eqnarray}
{\bf T}_{\rm damping}&=&-\left(\alpha+\frac{\mu_{\rm B}}{M_{\rm s} d}\frac{\Delta}{ev}\tau_\bot\right)(\partial_tm_x{\bf x}+\partial_tm_y{\bf y})\nonumber\\
&&+\alpha\partial_tm_z{\bf z}.
\end{eqnarray}
This anisotropic magnetic relaxation echoes the famous D'yakonov-Perel spin relaxation emerging in two dimensional electron gases \cite{Dyakonov1972}. In recent experimental reports \cite{mellnik2014,fan2014,Wang2015}, the electrical torque efficiency in TIs ranges from $(\mu_{\rm B}/\gamma M_{\rm s}d)\tau_\bot\approx 10^{-9}$ T$\cdot$m/V \cite{mellnik2014} to $\approx 10^{-7}$ T$\cdot$m/V \cite{fan2014,Wang2015}, depending on the temperature and thickness of the ferromagnet. Hence, adopting standard materials parameters ($\hbar v\sim 4$ eV$\cdot$\AA, $\Delta\sim1$ eV), we obtain a damping enhancement of $(\mu_{\rm B}/M_{\rm s} d)(\Delta/ev)\tau_\bot\approx$ 3$\times$10$^{-4}$ to 3$\times$10$^{-2}$, which is experimentally measurable. For the sake of comparison, the enhanced damping observed in Bi$_2$Se$_3$/CoFeB bilayers lies between $\sim 0.03$ and $\sim 0.12$, with wide variability from sample to sample \cite{Jamali2015}. 

\section{Magnetization switching by Dirac SOT}
\begin{figure}[b]
\includegraphics[width=8.4cm]{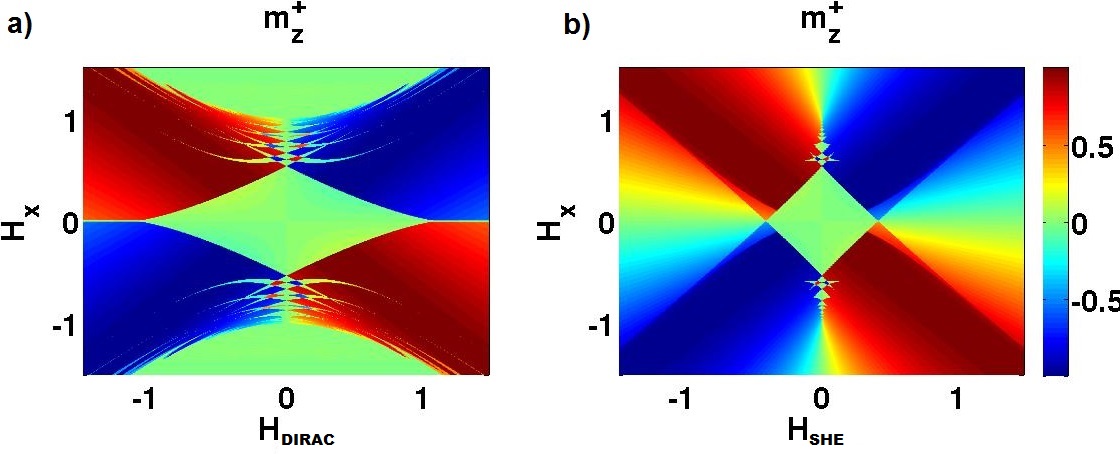}
\caption{\label{fig:2} (Color online) Calculated Switching Phase Diagram with an applied in-plane field along ${\bf x}$, for a current induced (a) Dirac torque and (b) spin Hall torque (retrieve results from Ref. \cite{sheliu2012})}
\end{figure}

We conclude this study by analyzing the impact of the Dirac damping-like torque on the magnetization reversal. In particular, we are interested in comparing the ability of the damping-like Dirac SOT ($\sim m_z{\bf m}\times{\bf E}$) with the damping-like SHE-induced SOT \cite{sheliu2012} ($\sim {\bf m}\times[({\bf z}\times{\bf E})\times{\bf m}]$) to switch the magnetization direction of a perpendicularly magnetized FM. We study the dynamics of the magnetization within the standard macrospin approximation and numerically solve the  LLG equation supplemented by SOT,
\begin{equation} \label{llg}
\partial_t {\bf m} = -\gamma{\bf m}\times{\bf H}_{\rm eff} + \alpha{\bf m}\times\partial_t{\bf m} + {\bf T}_{\rm dirac/she},
\end{equation}
where ${\bf H}_{\rm eff}$ is the effective field incorporating the demagnetizing field and/or an external applied magnetic field while the last term, $ {\bf T}_{\rm dirac/she}$, represents the (Dirac or SHE-induced) damping-like SOT. In the configuration we adopt, the current is driven along ${\bf x}$ and the magnetic anisotropy is along ${\bf z}$. The Dirac damping-like SOT is therefore ${\bf T}_{\rm dirac}=\gamma H_{\rm dir}m_z{\bf m}\times{\bf x}$, while the SHE-induced SOT is ${\bf T}_{\rm she}=\gamma H_{\rm she}{\bf m}\times({\bf y}\times{\bf m})$, $H_{\rm dir/she}$ being the strength of the torque. Solving the LLG equation, Eq. \eqref{llg}, when varying both the in-plane applied magnetic field $H_x{\bf x}$ and the SOT strength, one obtains the switching phase diagram of the macrospin as displayed in Fig. \ref{fig:2}. Notice that Fig. \ref{fig:2}(b) has been calculated previously \cite{sheliu2012} and is only reproduced here for comparison. 

Both diagrams display the same general shape: a central diamond-like region (green) denotes the bistable state where both $+{\bf z}$ and $-{\bf z}$ states are stable. This region is surrounded by four regions of monostable states (blue or red), when only $+{\bf z}$ or $-{\bf z}$ state is stable. Besides these general features, we observe two
major differences. First the horizontal extension of the central diamond is twice larger in the case of Dirac SOT than for SHE-induced SOT, which means that the SHE-induced SOT is twice as efficient as the Dirac SOT. This can be understood easily as the SHE-induced SOT has the form ${\bf m}\times({\bf y}\times{\bf m})=m_z{\bf m}\times{\bf x}-m_x{\bf m}\times{\bf z}$, while the Dirac damping-like SOT is simply $\sim m_z{\bf m}\times{\bf x}$. A second interesting aspect is the shape of the transitions between the monostable regions (red and blue). In the case of SHE-induced SOT, when varying the in-plane field $H_x$, there is a continuous variation between the two opposite stable states (from blue to red, and from red to blue). In contrast, in the case of Dirac SOT, the transition between the blue and red regions is much more abrupt, which is related to the vanishing of the Dirac damping-like SOT when the magnetization lies in the plane of the surface.

\section{Conclusion}

To summarize, we have analytically derived the electrically driven SOTs and charge pumping at the magnetic surface of a TI. While the field-like Dirac torque has the same geometrical form as the standard field-like Rashba torque, the damping-like Dirac torque presents a remarkable difference compared to the SHE-induced torque and vanishes when the magnetization lies in the plane of the surface. Furthermore, we uncover a strong angular dependence of the torque due to (i) the distortion of the band structure associated with the gap opening when the magnetization lies out-of-plane, and (ii) the presence of anisotropic spin relaxation.\par

 We note that a strong but opposite angular dependence of the torque has been experimentally reported in magnetically-doped topological insulators by Fan et al. \cite{fan2014}: in this experiment the magnitude of the torque is larger when the magnetization lies perpendicular to the plane of the surface. Another difference between Eq. (\ref{eq:torque}) and the experimental observations is that in Refs. \cite{fan2014,Fan2016}, the SOT is dominated by the damping-like component while in Eq. (\ref{eq:torque}), the field-like torque dominates. \par

The charge pumping induced by a time-varying magnetization presents similar features as it is the Onsager reciprocal of the SOTs. Interestingly, the pumped charge current in turns enhances the magnetic damping of the {\em in-plane} magnetization components. Although the magnitude of the enhanced damping calculated in the present work is consistent with the experimental observations \cite{Jamali2015}, one cannot exclude that other effects, such as SHE of the TI bulk states \cite{Sahin2015}, could also contribute to the spin-charge conversion process in these systems.

In conclusion, while the standard theory of magnetic TI surfaces derived in the present work can account for some of the features observed experimentally, some major discrepancies (in particular the angular dependence and the magnitude of the damping torque) cannot be explained. These limitations suggest that the coupling between the magnetic material and the TI surface \cite{scholz2012,Zhang2016}, as well as the contribution of bulk states \cite{Sahin2015} should be taken into account to model the experiments. 

\acknowledgments

PBN and AM were supported by the King Abdullah University of Science and Technology (KAUST). PBN thanks Z. T. Ndiaye for the fruitful discussions and valuable technical support, and AM acknowledges inspiring discussions with H. Yang, E. Y. Tsymbal and J. Zhang. EAK was supported by NSF CAREER award DMR-0955822 and by the Cornell Center for Materials Research with funding from the NSF MRSEC program (DMR-1120296). MHF acknowledges support from the Swiss Society of Friends of the Weizmann Institute of Science.\par

\par

\appendix*

\section{Integration of $\delta {\bf S}^{\rm II}$ when $\varepsilon_{\rm F}>\Delta$ }
In this appendix, we further clarify the significance of the Fermi sea contribution to the non-equilibrium spin density, $\delta {\bf S}^{\rm II}$, given by Eq. (\ref{kubostreda1}). Let us demonstrate that this contribution vanishes in the metallic regime. Such a demonstration has been carried out by Sinitsyn et al. \cite{sinitsyn2007} for the anomalous Hall effect and we now explicitly extend it to the non-equilibrium spin density. We can notice straightaway that $\delta {\bf S}^{\rm II}$ is, by construction, even in scattering time $1/\tau$ [Eq. (\ref{kubostreda1}) only involves terms like $\sim G^R_\varepsilon G^R_\varepsilon$ and $\sim G^A_\varepsilon G^A_\varepsilon$]. Therefore, in the limit of long relaxation time, $\delta {\bf S}^{\rm II}\approx \delta {\bf S}^{\rm II}_{\rm int}+O(1/\tau^2)$, where $\delta {\bf S}^{\rm II}_{\rm int}$ is the intrinsic contribution in the absence of disorder. The higher-order contributions lie beyond the scope of our study since we only search for terms $\sim \tau$ and $\sim 1+O(1/\tau)$ [see Eq. (\ref{sden})]. Hence, our aim is to demonstrate that the intrinsic contribution, $\delta {\bf S}^{\rm II}_{\rm int}$, vanishes.\par

The strategy is to write down Eq. (\ref{kubostreda1}) in the chiral basis $\left\{\ket{u_+^{\bf k}}, \ket{u_-^{\bf k}}\right\}$ given in Eq. (\ref{eq:chiralbasis}). In this basis, the retarded (advanced) Green's function read $\hat G^{R(A)}_\varepsilon=\sum_s\frac{\ket{u_s^{\bf k}}\bra{u_s^{\bf k}}}{\varepsilon-\varepsilon_{\bf k}^s\pm i0^+}$. Let us now decompose the energy integral into positive and negative energy regions
\begin{eqnarray} \label{sig-}
\delta {\bf S}^{\rm II-}_{\rm int}&=&\frac{\hbar}{2\pi}\Re\int_{-\infty}^{0} d\varepsilon  f(\varepsilon){\rm tr}\{ \cdots \} ,\\
\delta {\bf S}^{\rm II+}_{\rm int}&=&\frac{\hbar}{2\pi}\Re\int_{0}^{+\infty} d\varepsilon  f(\varepsilon){\rm tr}\{ \cdots \}. \label{sig+}
\end{eqnarray}
In the different regions, the unperturbed Green's function reads 
\begin{eqnarray}
\hat G^{R(A)}_{\varepsilon<0}&=&\frac{\ket{u_+^{\bf k}}\bra{u_+^{\bf k}}}{\varepsilon-\varepsilon_{\bf k}^+}+\frac{\ket{u_-^{\bf k}}\bra{u_-^{\bf k}}}{\varepsilon-\varepsilon_{\bf k}^-\pm i0^+},\\
\hat G^{R(A)}_{\varepsilon>0}&=&\frac{\ket{u_+^{\bf k}}\bra{u_+^{\bf k}}}{\varepsilon-\varepsilon_{\bf k}^+\pm i0^+}+\frac{\ket{u_-^{\bf k}}\bra{u_-^{\bf k}}}{\varepsilon-\varepsilon_{\bf k}^-}.
\end{eqnarray}
Then, Eqs. (\ref{sig-}) and (\ref{sig+}) can be rewritten 
\begin{widetext}
\begin{eqnarray} \label{sig-1}
\delta {\bf S}^{\rm II-}_{\rm int}&=&\frac{\hbar}{2\pi}\Re\int_{-\infty}^{0} d\varepsilon  f(\varepsilon)\int\frac{d^2{\bf k}}{(2\pi)^2}{\bf F}^{+-}_{\bf k}\left[\frac{1}{\varepsilon-\varepsilon_{\bf k}^-+i0}\partial_\varepsilon\frac{1}{\varepsilon-\varepsilon_{\bf k}^+}-\partial_\varepsilon\frac{1}{\varepsilon-\varepsilon_{\bf k}^-+i0}\frac{1}{\varepsilon-\varepsilon_{\bf k}^+}\right] ,\\
\delta {\bf S}^{\rm II+}_{\rm int}&=&\frac{\hbar}{2\pi}\Re\int_{0}^{+\infty} d\varepsilon  f(\varepsilon)\int\frac{d^2{\bf k}}{(2\pi)^2}{\bf F}^{+-}_{\bf k}\left[\frac{1}{\varepsilon-\varepsilon_{\bf k}^-}\partial_\varepsilon\frac{1}{\varepsilon-\varepsilon_{\bf k}^++i0}-\partial_\varepsilon\frac{1}{\varepsilon-\varepsilon_{\bf k}^-}\frac{1}{\varepsilon-\varepsilon_{\bf k}^++i0}\right] \label{sig+1},
\end{eqnarray}
\end{widetext}
where ${\bf F}^{+-}_{\bf k}=\langle u_{\bf k}^+|\hat{\bm\sigma}|u_{\bf k}^-\rangle\langle u_{\bf k}^-|(\hat{\bf v}\cdot e{\bf E})|u_{\bf k}^+\rangle$. Let us now transform Eqs. (\ref{sig-1}) and (\ref{sig+1}) so that the $\delta$-functions appear explicitly. The second part of Eq. (\ref{sig-1}), $\sim\partial_\varepsilon\frac{1}{\varepsilon-\varepsilon_{\bf k}^-+i0}$ can be manipulated by performing integration by part. This way, the energy derivative is distributed over $1/(\varepsilon-\varepsilon_{\bf k}^+)$ and $f(\varepsilon)$. Since $f(\varepsilon)$ is constant in the range $]-\infty,0]$, the contribution $\sim\partial_\varepsilon f(\varepsilon)$ vanishes. We perform the same integration by part on the term $\sim\partial_\varepsilon\frac{1}{\varepsilon-\varepsilon_{\bf k}^++i0}$ in Eq. (\ref{sig+1}), but now $\partial_\varepsilon f(\varepsilon)$ does not vanish at the Fermi energy. Overall, we obtain
\begin{eqnarray}
\delta {\bf S}^{\rm II-}_{\rm int}&=&-\hbar\Im\int\frac{d^2{\bf k}}{(2\pi)^2}\frac{{\bf F}^{+-}_{\bf k}}{(\varepsilon_{\bf k}^+-\varepsilon_{\bf k}^-)^2}\label{sii-}\\
\delta {\bf S}^{\rm II+(a)}_{\rm int}&=&-\hbar\Im\int\frac{d^2{\bf k}}{(2\pi)^2}f(\varepsilon_{\bf k}^+)\frac{{\bf F}^{+-}_{\bf k}}{(\varepsilon_{\bf k}^+-\varepsilon_{\bf k}^-)^2}\label{sii+a}\\
\delta {\bf S}^{\rm II+(b)}_{\rm int}&=&-\frac{\hbar}{2}\Im\int\frac{d^2{\bf k}}{(2\pi)^2}{\bf F}^{+-}_{\bf k}\frac{\delta(\varepsilon_{\rm F}-\varepsilon_{\bf k}^+)}{\varepsilon_{\rm F}-\varepsilon_{\bf k}^-}.\label{sii+b}
\end{eqnarray}
In Eq. (\ref{sii-}), the integration over $k$ lies in the range $[0,+\infty[$, while in Eq. (\ref{sii+a}) it runs over $[0,k_{\rm F}]$, where $k_{\rm F}$ is the solution of $\varepsilon_{\bf k}^+=\varepsilon_{\rm F}$ and depends on the angle $\phi_k$. Therefore, one can rewrite these expression explicitly
\begin{widetext}
\begin{eqnarray}
\delta {\bf S}^{\rm II-}_{\rm int}&=&\frac{\hbar v}{2}\int_0^{2\pi}\int_0^{+\infty}\frac{d\phi_kkdk}{(2\pi)^2}\frac{1}{\varepsilon_{\bf k}^3}\left(\Delta(\cos\theta e{\bf E}-({\bf m}\cdot e{\bf E}){\bf z})+\hbar v[({\bf z}\times{\bf k})\cdot e{\bf E}]{\bf z}\right)\\
\delta {\bf S}^{\rm II+(a)}_{\rm int}&=&-\delta{\bf S}^{\rm II-}_{\rm int}+\frac{\hbar v}{2}\int_0^{2\pi}\int_{k_{\rm F}}^{+\infty}\frac{d\phi_kkdk}{(2\pi)^2}\frac{1}{\varepsilon_{\bf k}^3}\left(\Delta(\cos\theta e{\bf E}-({\bf m}\cdot e{\bf E}){\bf z})+\hbar v[({\bf z}\times{\bf k})\cdot e{\bf E}]{\bf z}\right)\\
\delta{\bf S}^{\rm II+(b)}_{\rm int}&=&\frac{\hbar v}{2\varepsilon_{\rm F}}\int_0^{2\pi}\int_0^{+\infty}\frac{d\phi_kkdk}{(2\pi)^2}\left(\Delta(\cos\theta e{\bf E}-({\bf m}\cdot e{\bf E}){\bf z})+\hbar v[({\bf z}\times{\bf k})\cdot e{\bf E}]{\bf z})\delta(\varepsilon_{\rm F}-\varepsilon_{\bf k}^+\right).\end{eqnarray}
\end{widetext}
Hence, it is sufficient to calculate $\delta{\bf S}^{\rm II+(a)}_{\rm int}+\delta{\bf S}^{\rm II-}_{\rm int}$ and $\delta{\bf S}^{\rm II+(b)}_{\rm int}$. After some algebra, we get
\begin{eqnarray}
\delta{\bf S}^{\rm II+(a)}_{\rm int}+\delta{\bf S}^{\rm II-}_{\rm int}=\frac{\Delta}{4\pi}\frac{1}{\hbar v\varepsilon_{\rm F}}\cos\theta e{\bf E},\\
\delta{\bf S}^{\rm II+(b)}_{\rm int}=-\frac{\Delta}{4\pi}\frac{1}{\hbar v\varepsilon_{\rm F}}\cos\theta e{\bf E},
\end{eqnarray}
and then, $\delta{\bf S}^{\rm II}_{\rm int}=\delta{\bf S}^{\rm II+(a)}_{\rm int}+\delta{\bf S}^{\rm II-}_{\rm int}+\delta{\bf S}^{\rm II+(b)}_{\rm int}=0$. Consequently, the Fermi sea contribution to the non-equilibrium electrically induced spin density vanishes in the metallic limit, within the weak scattering regime.

\end{document}